\documentclass[11pt,twoside]{article}
\usepackage{asp2004}
\usepackage{psfig}
\usepackage{epsf}
\usepackage{graphics}
\usepackage{lscape}
\def\edcomment#1{\iffalse\marginpar{\raggedright\sl#1\/}\else\relax\fi}
\reversemarginpar

\begin{document}
\title{Evolution of the Most Massive Stars}
\author{Andr\'e  Maeder, Georges Meynet \& Raphael Hirschi}
\affil{Geneva Observatory, CH-1290 Sauverny, Switzerland}

\begin{abstract}
We discuss the physics of the  $\Omega \Gamma$-- Limit, i.e. when the star is unbound
as a result of {\emph{both}} rotation and radiation pressure. We suggest that
the  $\Omega \Gamma$-- Limit is what makes the Humphreys--Davidson Limit.
Stellar filiations  are discussed, with an emphasis on the final stages, in particular on the final masses, their angular momentum and the chemical yields
with account of rotation. A possible relation between WO stars and GRB is emphasized.
\end{abstract}
\thispagestyle{plain}

\section{Introduction}
Massive stars are at the crossroads of major subjects  in Astrophysics. They
have a  key influence on the spectral  and chemical evolution of galaxies.
Their contributions to the  spectrum of starbursts is
observed up to large redshifts. They  are also the essential
components of the metallicity $Z=0$ populations, the progenitors of WR stars,
 supernovae and  gamma ray bursts (GRB).
 
\section{The $\Omega \Gamma$-- Limit: physics of the interaction of mass loss and rotation}

The physics and evolution of massive  stars is dominated by 
mass loss and by  rotational mixing. At the origin of 
these two effects, we find the large ratio $T/\rho$ of
temperature to density in massive stars. This   enhances the ratio of radiation 
to gas pressure, which goes like $T^3/\rho$ and  
favours  stellar winds. A high $T/\rho$ enhances 
rotational mixing, since mixing by shear turbulence
scales as the thermal diffusivity $K=4acT^3/3 c_{\mathrm{P}} \,  \kappa \, \rho^2$.
 Also,  the velocity of
meridional circulation scales as the ratio $L/M$ of the luminosity to mass.

Let us recall  that there are 3 main effects of   rotation. 1) Structural effects due 
to the centrifugal force: these are small in the interior, but 
may produce a large distorsion at the  surface. 
2) Rotational mixing: this produces internal transports of chemical elements and angular momentum.
Mixing by shears is  most efficient for the transport of chemical elements
 \citep{Z92,MMVII}, while meridional circulation dominates the transport of angular momentum. 
3) Enhancement of mass loss  by rotation: an important case is the 
$\Omega \Gamma$-- Limit, where the star reaches break--up as a result
 of both high radiation pressure and rotation.
 \begin{figure}[!ht]
\plottwo{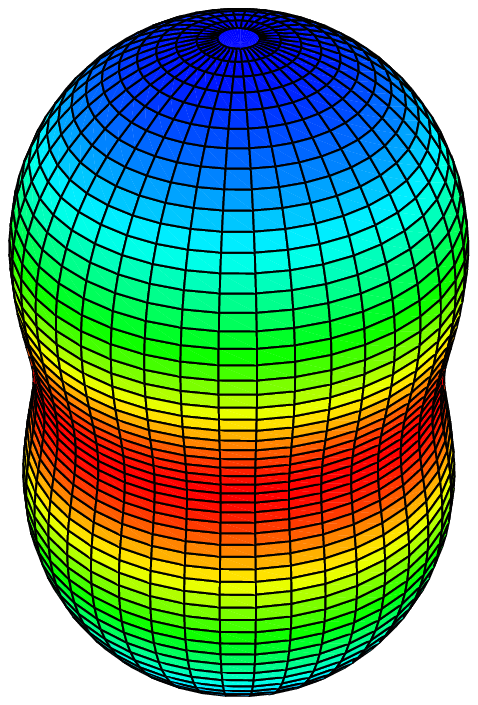}{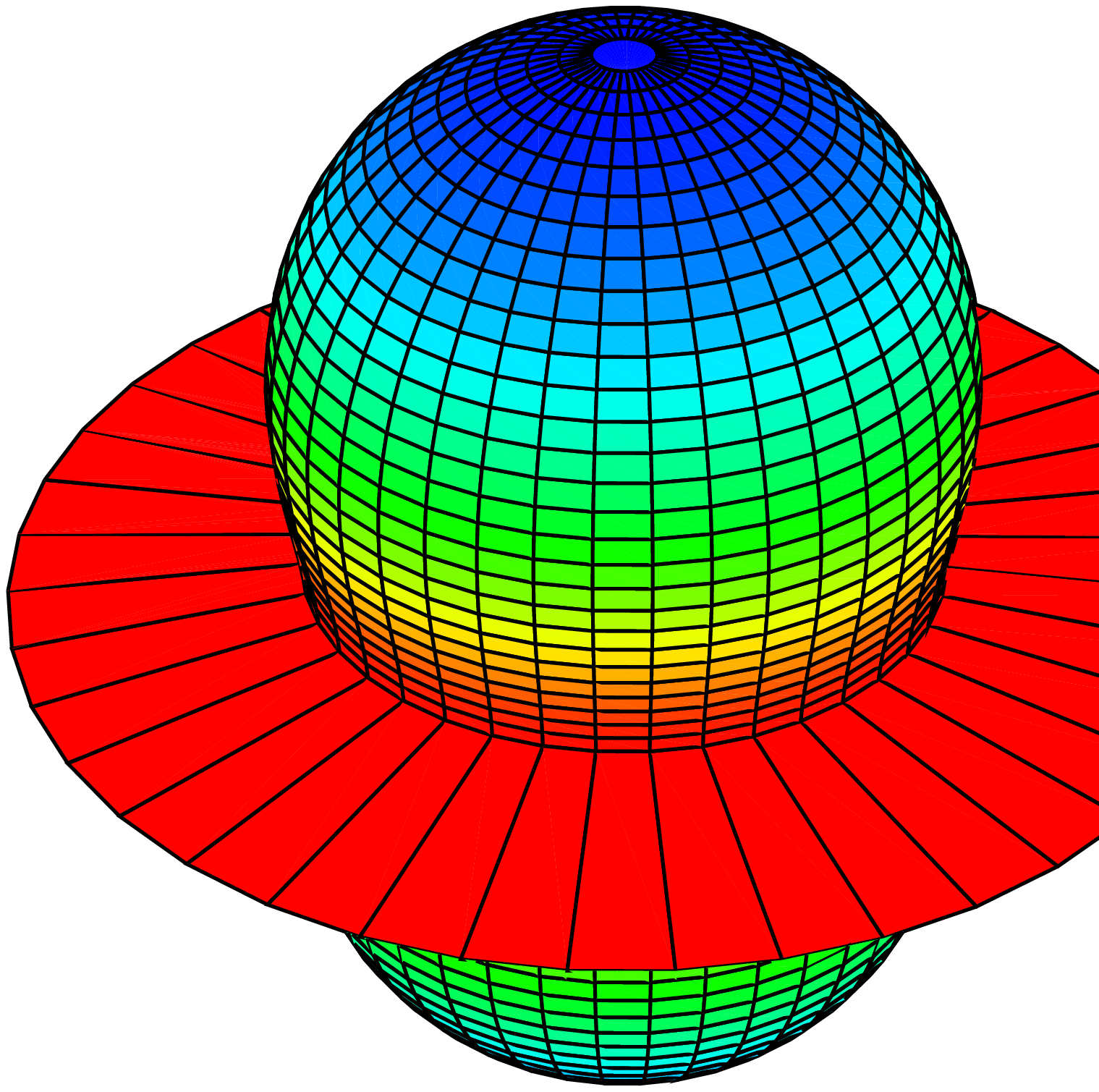}
\caption{Left: Iso-mass loss distribution for a 120 $M_{\odot}$ 
star with $\log{\frac{L}{L_{\odot}}=6.0}$
and  T$_{\mathrm{eff}}=30 000$ K rotating at a fraction 0.80 of break--up velocity. Right:
The same  with T$_{\mathrm{eff}}=25 000$ K.}
\end{figure}%Fig. 1
The von Zeipel theorem states that the local radiative flux $F$ of a rotating star 
 is proportional to the effective gravity $g_{\mathrm{eff}}$. Thus, there is 
  a much larger flux and a higher
$T_{\mathrm{eff}}$ at the pole than at the equator  \citep{MMI}.  This latitudinal dependence of
$T_{\mathrm{eff}}$ leads to asymmetric mass loss and also to enhanced 
average mass loss rates.

On a rotating star, one must consider 
the flux $F(\vartheta)$ at a given colatitude $\vartheta$ as given by von Zeipel's theorem,
$F(\vartheta)  =  - \frac{L(P)}{4 \pi GM_{\star}}
g_{\rm{eff}} [1 + \zeta(\vartheta)]$.
The term $\zeta(\vartheta)$  is in general negligible. 
 The Eddington factor is a local quantity $\Gamma_{\Omega}(\vartheta)$ depending on $\vartheta$
 and rotation.
 We define it  as
the ratio of the local flux $F(\vartheta)$ 
given by the von Zeipel theorem to the maximum possible local flux,
which is $F_{\mathrm{lim}}(\vartheta) = - \frac{c}{\kappa(\vartheta)}  
g_{\mathrm{eff}}(\vartheta)$. Thus, one has
\begin{eqnarray}
\Gamma_{\Omega}(\vartheta) =
\frac{F(\vartheta)}{F_{\mathrm{lim}}(\vartheta)}=
\frac{ \kappa (\vartheta) \; L(P)}{4 \pi 
cGM \left( 1 - \frac{\Omega^2}{2 \pi G \rho_{\rm{m}}}  \right) } \; ,
\end{eqnarray}
\noindent 
where the opacity $\kappa(\vartheta)$ depends on the colatitude $\vartheta$, since $T_{\mathrm{eff}}$
itself depends on $\vartheta$. For electron scattering, $\kappa$ is constant 
and $\Gamma_{\Omega}(\vartheta)$ is the same at all latitudes, i.e. 
$\Gamma_{\Omega}= \Gamma / \left( 1 - \frac{\Omega^2}{2 \pi G \rho_{\rm{m}}} \right)$, where 
$\Gamma$ is the usual expression.
 Eq.~(1) shows 
that the maximum luminosity of a rotating star is reduced by rotation.
We see that the  dependences 
of $F(\vartheta)$ and of $F_{\mathrm{lim}}(\vartheta)$ with respect to $g_{\mathrm{eff}}$
have cancelled each other in the expression of $ \Gamma_{\Omega}(\vartheta)$. 
Thus, if the limit  $\Gamma_{\Omega}(\vartheta) = 1 $ 
happens to be met  at the equator, it is not because 
$g_{\mathrm{eff}}$ is the lowest there, but because the 
opacity is the highest! 
\begin{figure}[!ht]
\plotone{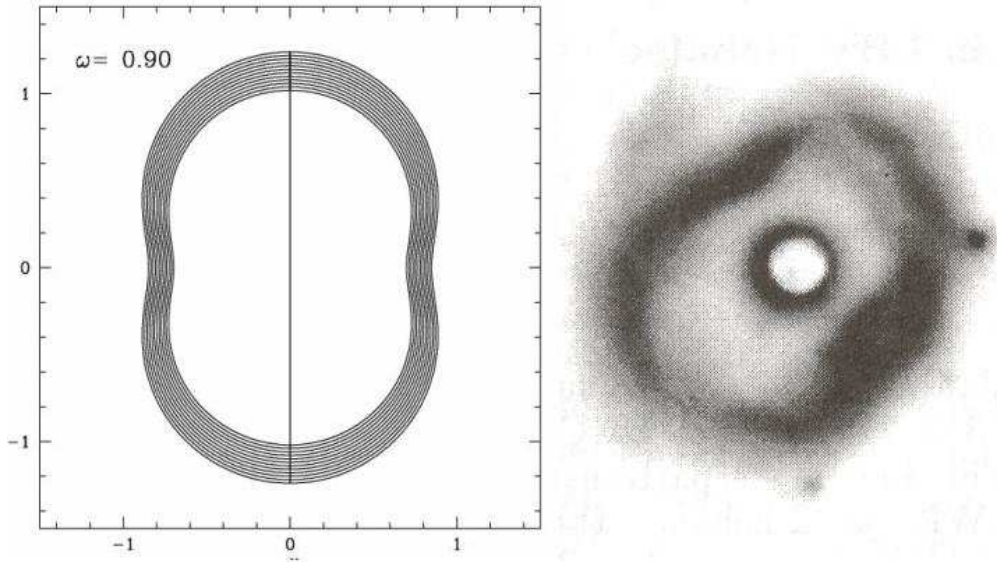}
\caption{Left: simulation of a short shell ejection by a massive star with anisotropic mass loss 
as in Fig.~1 left. Right:  the nebulae around  AG Carinae \citep{Nota97}.}
\end{figure}
The theory of radiative winds applied to a  rotating star 
leads to an  expression  of the mass flux as a function of 
colatitude  \citep{Maeder99}.
Figs.~1 left  and  right  \citep{MaeDes00} illustrate the distribution of the mass loss rates
around a massive star of 120 $M_{\odot}$ for two different $T_{\mathrm{eff}}$.
For a star hot enough to have electron scattering opacity 
 from pole to equator, the iso--mass loss curve has a peanut--like 
shape (Fig.~1 left). This  results from the fact that the pole is hotter 
(``$g_{\mathrm{eff}}$--effect'').
For a rotating star with a lower $T_{\mathrm{eff}}$ (Fig.~1 right),  a bistability limit
i.e. a steep increase of the opacity \citep{Lamers95} may occur somewhere between the 
pole and the equator. This ``opacity--effect'' produces an equatorial enhancement
of the mass loss (Fig.~1 right). In Fig.~2 left, we show the model of a short shell
ejection with mass loss corresponding to the peanut--shape. The  corresponding 
 image of AG Carinae \citep{Nota97} is shown in Fig.~2 right.

The anisotropies of mass loss influence the loss of angular
momentum. Polar mass loss removes mass but relatively little 
angular momentum. This has a great incidence on the evolution of the most massive stars
with high rotation \citep{MIX}. 
The mass loss rate $\dot{M} (\Omega)$ of a rotating star 
compared to that $\dot{M} (0)$ of a non--rotating star at the same location in the HR
diagram is given by \citep{MMVI},
\begin{equation}
\frac{\dot{M} (\Omega)} {\dot{M} (0)} \simeq
\frac{\left( 1  -\Gamma\right)
^{\frac{1}{\alpha} - 1}}
{\left[ 1 - 
\frac{4}{9} (\frac{v}{v_{\mathrm{crit, 1}}})^2-\Gamma \right]
^{\frac{1}{\alpha} - 1}} \; ,
\end{equation}
\noindent
where  $\alpha$ is a force multiplier \citep{Lamers95}. 
For a 10 $M_{\odot}$ star on the MS, 
$\frac{\dot{M} (\Omega)} {\dot{M} (0)}$ may reach 1.5. 
For the most luminous stars  which have a value 
$\Gamma$ close to 1.0, this  may be 
orders of magnitude, when the star
reaches break-up at the $\Omega \Gamma$--Limit.
 %For blue and red supergiants, the increase could also be large, however supergiants
%do not rotate fast in general. 

Often, the critical velocity in a rotating star is written as
$v^2_{\rm{crit}} = \frac{GM}{R} (1-\Gamma)$. This expression is not correct,
as it applies only to uniformly bright stars. 
Indeed, the critical velocity of a rotating star is given by the zero of the equation
expressing the total gravity 
$\vec{g_{\mathrm{tot}}} = 
 \vec{g_{\mathrm{grav}}} + \vec{g_{\mathrm{rot}}} + \vec{g_{\mathrm{rad}}}=
  \vec{g_{\mathrm{eff}}} \left[ 1 - \Gamma_{\Omega}(\vartheta) \right]$.
This equation has two roots \citep{MMVI}. The first that is met determines the critical 
velocity. The first root is as usual  $v_{\mathrm{crit, 1}} = 
\left( \frac{2}{3} \frac{GM}{R_{\mathrm{pb}}} \right)^{\frac{1}{2}}$, where
$R_{\mathrm{pb}}$ is the polar radius at break--up.
The second root  $v_{\mathrm{crit, 2}}$ applies to Eddington factors bigger than 0.639. It is
equal to 0.85, 0.69, 0.48, 0.35, 0.22, 0 times $v_{\mathrm{crit, 1}}$ for $\Gamma=$
0.70, 0.80, 0.90, 0.95, 0.98 and 1.00 respectively.

\section{Close to the Eddington Limit, evolution unavoidably reaches the $\Omega\Gamma$--Limit}

On the whole, we may distinguish 3 critical cases, where the outer  layers
escape: 1.-- The usual
$\Gamma$--Limit, when radiation effects largely dominate over rotation; 
this is the classical case. 2.-- The $\Omega$--Limit, when
rotation effects, rather than radiative effects, are determining break--up. This is relevant
for Be-stars, which are far enough from $\Gamma$ equal to unity. 
3.-- The $\Omega \Gamma$--Limit, when both rotation and radiation are important.  
This is the case which  applies to the most massive stars. 
\emph{Even for a rather  small initial rotation velocity of, say, $\geq 50$ km/s, a star with a 
high  $\Gamma$ will reach  critical velocity during its MS evolution}. This occurs because
the rotation velocity increases during MS evolution (see Fig.~3), while
the critical velocity, which  is 
given by $v_{\mathrm{crit, 2}}$,  decreases. The mass loss rates  increase
strongly (cf. Eq.~2), when the critical $\Omega \Gamma$--Limit is approached.
Such a situation has already been considered  \citep{La97},
however for the case of solid body rotation, while here 
the evolution of the angular momentum is followed with appropriate critical velocity and mass loss expressions.
\begin{figure}[!ht]
\plottwo{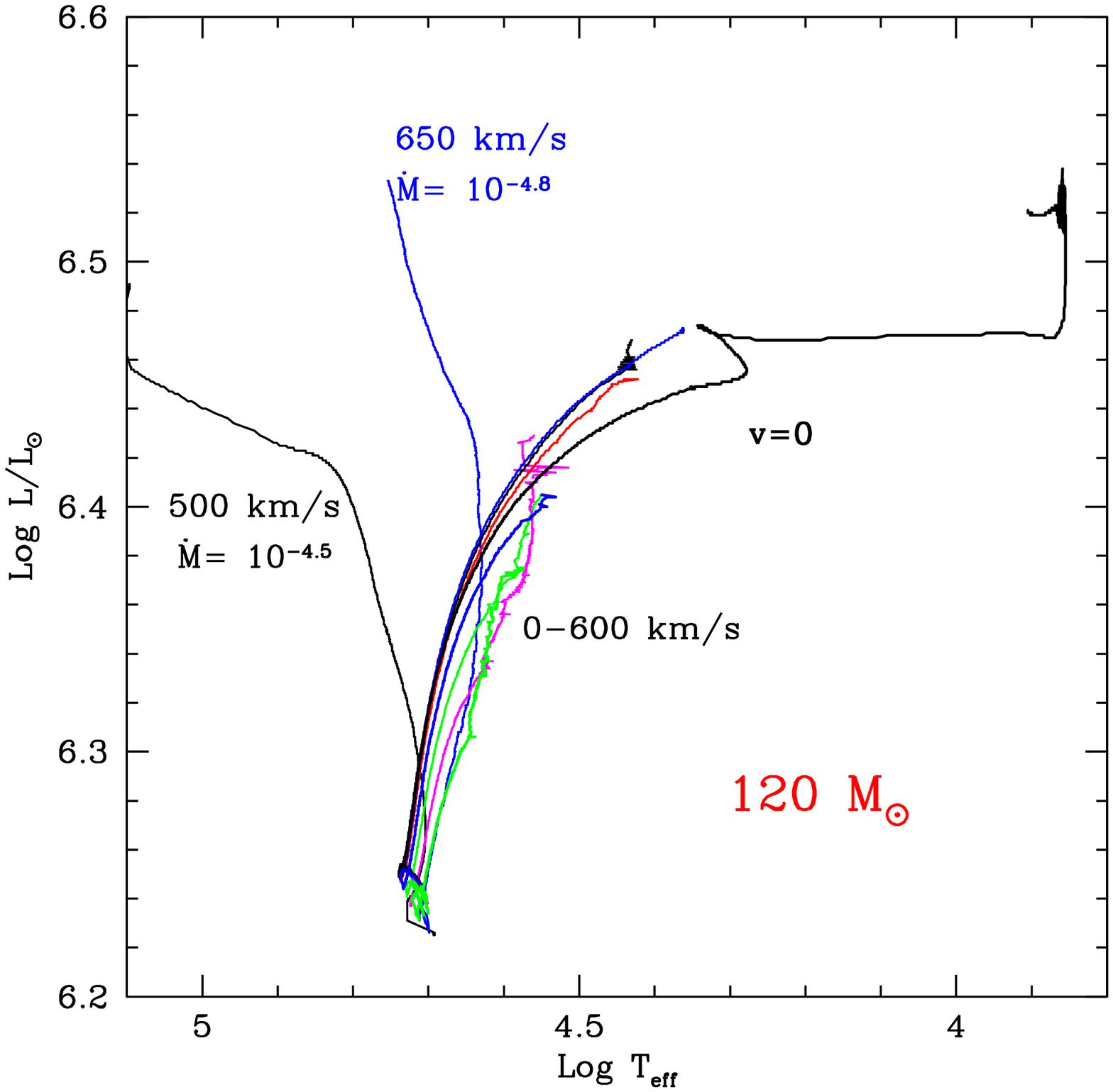}{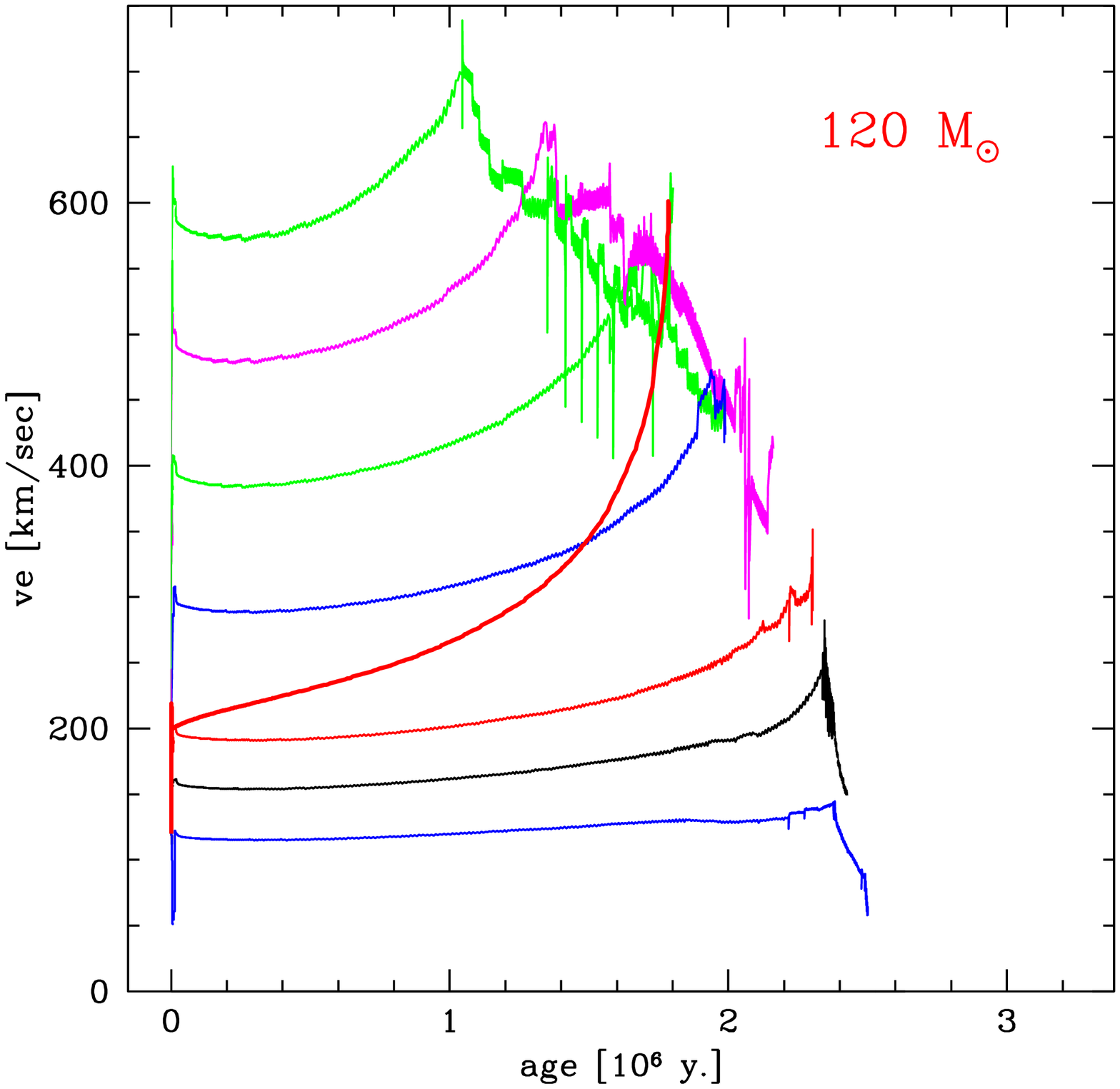}
\caption{Left: Evolution of a 120 $M_{\odot}$ star with different mass loss rates and initial rotational velocities  $v_{\mathrm{ini}}$, see text. Right: Evolution of  the rotational velocities
of models with different $v_{\mathrm{ini}}$. The envelope of the various curves  
is the location
where the stars reach their critical velocities. The models with a lower $v_{\mathrm{ini}}$ reach the critical velocity later. A model with solid body rotation
with $v_{\mathrm{ini}}=200$ km/s is also shown. It reaches the critical velocity faster. }
\end{figure}

The evolution of star reaching the $\Omega \Gamma$--Limit is illustrated in Fig.~3.
On the left, it shows different cases of tracks in the HR diagram.
The track  with $v_{\mathrm{ini}}= 500$ km/s and an average mass loss rate 
$\dot{M}$ of $10^{-4.5}$ $M_{\odot}\cdot yr^{-1}$
is calculated according to recent $\dot{M}$--rates \citep{Vink}. 
With these rates,  even the zero rotation track would turn bluewards!
 Thus,  at $120 M_{\odot}$, these rates  are likely too high 
 (at least for producing a star like  
 $\eta$ Carinae). With rates a factor of 2-3 lower, a bluewards evolution is obtained only
 for a very fast rotation, such as 650 km/s. For initial rotation $v_{\mathrm{ini}}= 600$ km/s
 or lower, redward evolution is obtained, with tracks having a different MS
 extension (for low rotation, the tracks are higher and extend further in the HR diagram than
 the track for zero rotation; for fast rotation, the extension is shorter as the result of the lower  opacity due to higher helium enrichment). 
 All tracks with $v_{\mathrm{ini}} \geq$  50 km/s reach  the  $\Omega \Gamma$--Limit
 with break--up velocity, thus experiencing phases of extreme mass loss.

  Fig.~3 right illustrates the corresponding behaviour of the rotational velocities.
  We see that the velocities increase during evolution.  The faster 
  $v_{\mathrm{ini}}$, the earlier the $\Omega \Gamma$--Limit is reached (envelope
  of the curves).  We  see that solid body rotation,
   which is the extreme case of coupling of the angular momentum, leads the
  star very promptly to the $\Omega \Gamma$--Limit.
   The milder coupling realized by meridional circulation
   leads the star to the $\Omega \Gamma$--Limit more slowly.
  We conclude that the most massive stars reach an $\Omega \Gamma$--Limit,
  which is different for each star according to its rotational velocity. The  
  $\Omega \Gamma$--Limit is the physical reason for the observed Humphreys--Davidson 
  Limit \citep{HD79}.

\section{What occurs at the $\Omega \Gamma$--Limit ?}

 \begin{figure}[!ht]
\plottwo{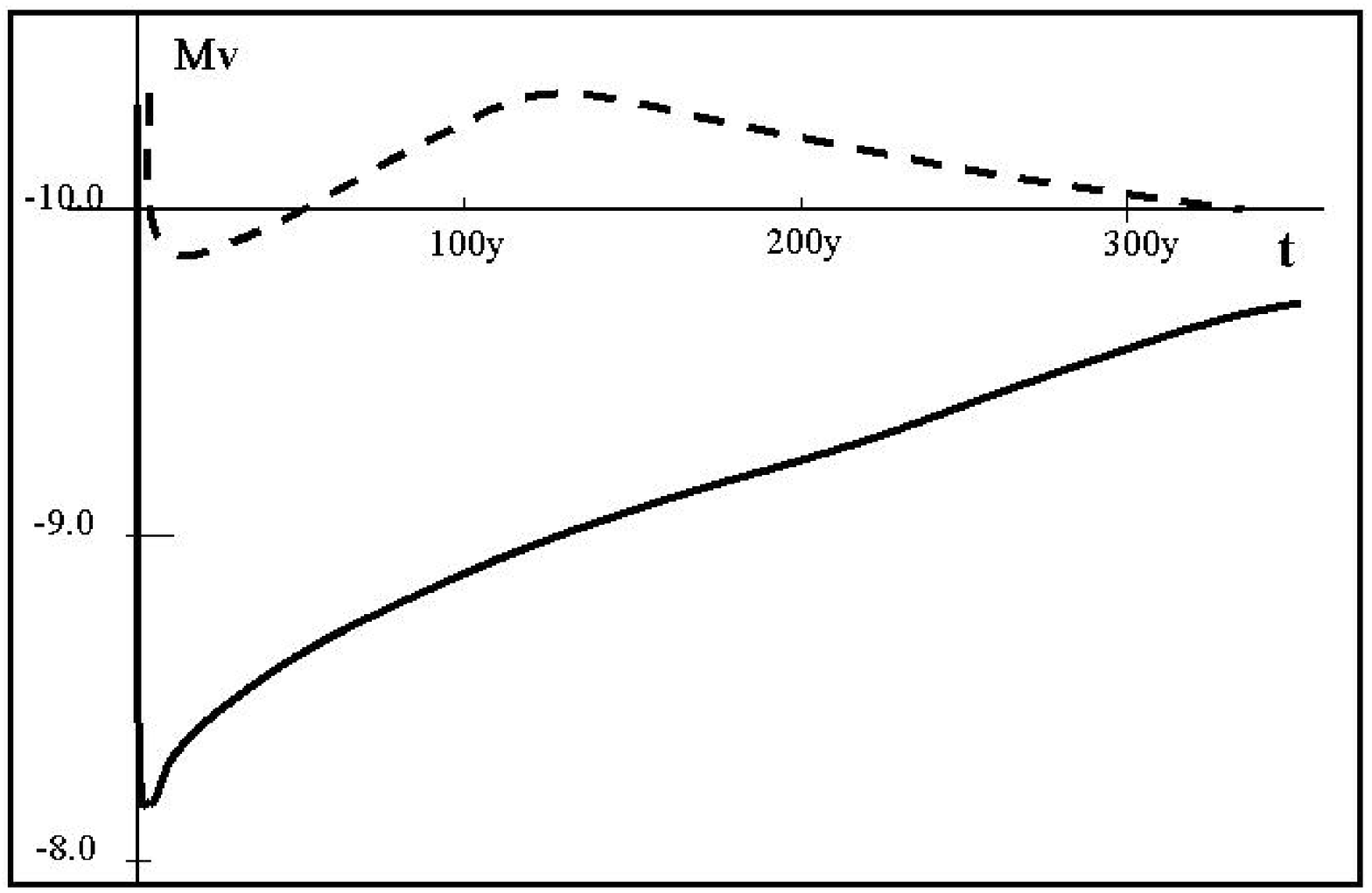}{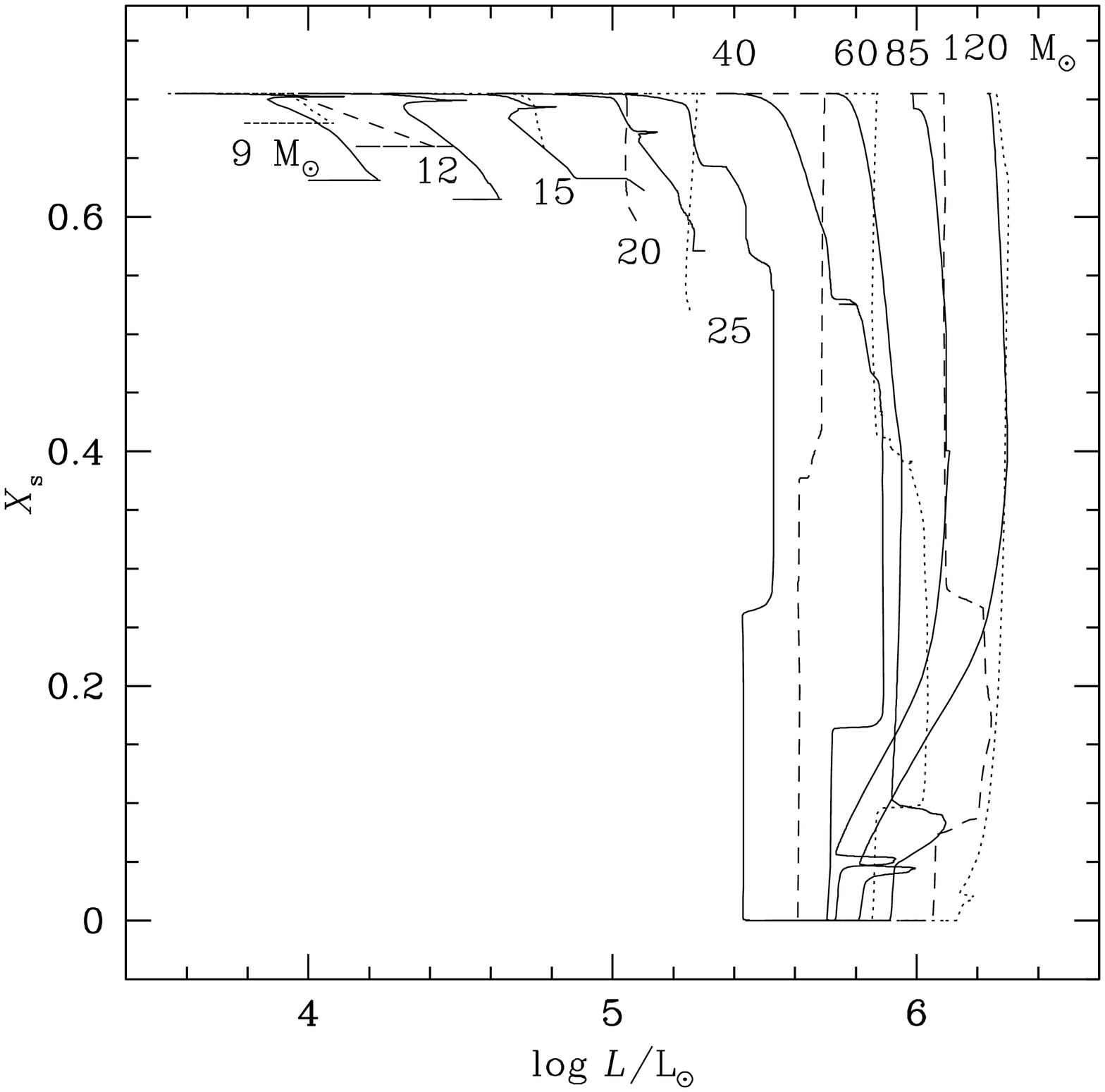}
\caption{Left: The broken line shows the  V--light curve for a star with
$\log T_{\mathrm{eff}}=3.8$ experiencing a mass ejection of $1.0 M_{\odot}$, 
the continuous line shows the same for an ejection of 3 $M_{\odot}$, the last part of this
curve would be similar to the broken line. Right: Evolution of the massive stars  in the 
plot of the surface H-content $X_{\mathrm{s}}$ vs. luminosity.}
\end{figure}
  For now, hydrodynamical  models of  LBV  outbursts do not exist, nevertheless
  we can make a few remarks on their properties:\\
-1. Likely, the most massive stars reach the
 $\Omega \Gamma$--Limit on their redwards track as illustrated in Fig.~3 left.
 The slightly less massive stars, firstly evolving to the red, may likely reach the 
 Limit on their bluewards tracks, because as a result of their mass loss,
  their  L/M is larger when they go back to the blue.
   Also,  due to the coupling
 of angular momentum  by convection, they may rotate faster when evolving bluewards
 \citep{La98}.\\
-2. As stated above, the $\dot{M}$--rates increase as given by Eq.~(2), when the
  $\Omega \Gamma$--Limit is approached. However, at the Limit, the $\dot{M}$--rates 
  are no longer given by the stellar wind theory. We think that the 
  $\dot{M}$--rates are such that the star just remains at the critical velocity, (or in  practice,
  at a fraction e.g. 0.98 or 0.99 of it). If the $\dot{M}$--rates would 
  be larger the star would soon become subcritical, if they would be smaller, 
  the star would rotate much above  critical.\\
-3. Some layers become supra-Eddington, which produces a density inversion
  and convective instability appears. Thus, even a B--type supergiant can have convection
  in its outer layers \citep{Mae80}.\\
-4. In the very massive stars, the thermal timescale may become smaller than the dynamical timescale in the outer layers. This means that the thermal structure
  of these layers can readjust during a dynamical event. Thus, the ionization front may move 
  inwards and allow a substantial mass fraction to participate to a dynamical event. This is
  the idea of the ``geyser model'' \citep{Mae92,HD94,Mae97}, which leads to an
  estimate of the amount of mass ejected  as a function of the luminosity. This has been 
  confirmed by IRAS observations  of dust in LBV nebulae \citep{Huts97}.\\
-5. As a result of the mass  ejected in the outburst, the star is shifted to the  blue in
  the HR diagram, then it slowly recovers its previous $T_{\mathrm{eff}}$. The luminosity keeps
  nearly constant, apart from the energy lost in the mechanical event.
   Simulations of sudden large mass ejection show several properties: - a) The extension of  blue shift in the HR diagram depends
  on the ejected mass  $\Delta M$; - b) The timescale of the blueshift is determined by the
  mass  loss rate in the outburst; 
  - c) The light curve in V magnitude (cf. Fig.~4 left) is mainly determined by the change 
  of the bolometric correction (see lightcurve of $\eta$ Car);
   - d) The recovery time, i.e. the time for the star to 
  reach its former equilibrium  depends on the ejected mass $\Delta M$.
  It is about 125, 350 and 750 yr for $\Delta M= 0.3, 1.0$ and 3.0 $M_{\odot}$. 
  
  \section{Filiations and further evolution of the massive stars}
  
\begin{figure}[!ht]
\plotfiddle{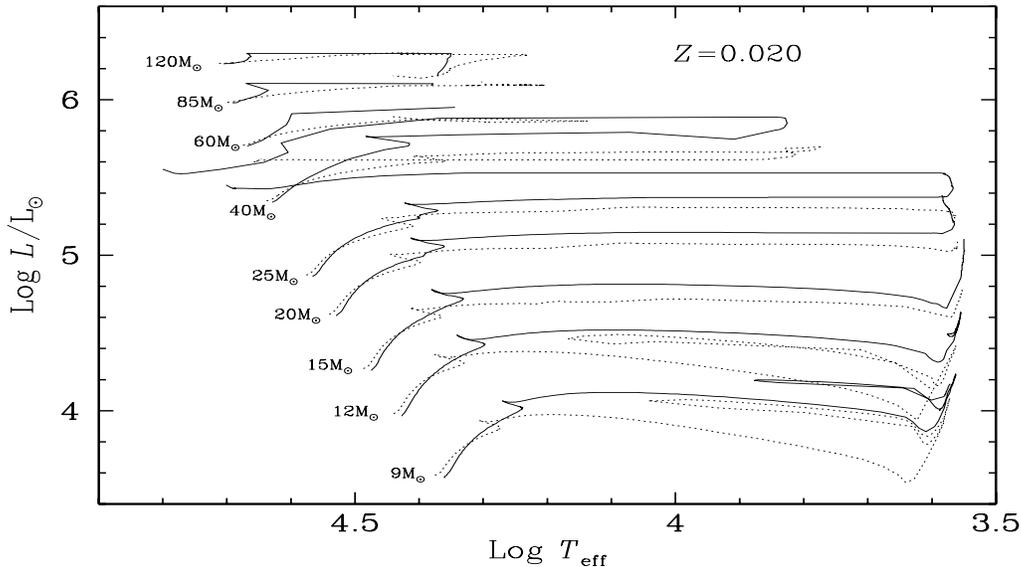}{7.0cm}{270}{50}{40}{-200}{230}
%\plotone{hrX.eps}
\caption{HR diagram of massive stars with metallicity Z=0.02}
\end{figure}
There are various evolutionary sequences: 
for the highest masses with mass loss, the stars may keep  to the blue. Then
in an interval of lower masses, the tracks have a certain extension to the
red and even in a lower mass interval the full extension to the red supergiants is 
present (Fig.~5 ). We may distinguish the following tentative filiations at solar $Z$:\\
\noindent
{\bf \underline{{$M >90  M_{\odot}$}}}:  O –- Of –- WNL –- (WNE) -– WCL –- WCE -– 
SN (Hypernova low Z  ?)\\
\noindent
{\bf \underline{{$60-90 \; M_{\odot}$}}}: O –- Of/WNL$<->$LBV -– WNL(H poor)-– WCL-E -– SN(SNIIn?)\\
\noindent
{\bf \underline{{$40-60 \; M_{\odot}$}}}: O –- BSG –-  LBV $<->$ WNL -–(WNE) -- WCL-E –- SN(SNIb) \\
\hspace*{5.9cm}  - WCL-E - WO – SN (SNIc) \\
\noindent
{\bf \underline{{$30-40 \; M_{\odot}$}}}:  O –- BSG –- RSG  --  WNE –- WCE -– SN(SNIb)\\
\hspace*{4.0cm}                        OH/IR $<->$ LBV ? \\
\noindent
{\bf \underline{{$25-30 \; M_{\odot}$}}}: O -–(BSG)–-  RSG  -- BSG (blue loop) -- RSG  -- SN(SNIIb, SNIIL)\\
\noindent
{\bf \underline{{$10-25 \; M_{\odot}$}}}: O –-  RSG -– (Cepheid loop, $M < 15 \; M_{\odot}$) – RSG -- 
SN (SNIIL, SNIIp)\\  

\noindent
The sign $<->$ means back and forth motions between the two  stages. The limits 
between the various scenarios  depend on metallicity $Z$ and rotation.
  The various types of supenovae are tentatively indicated. 

\begin{figure}[!ht]
\plottwo{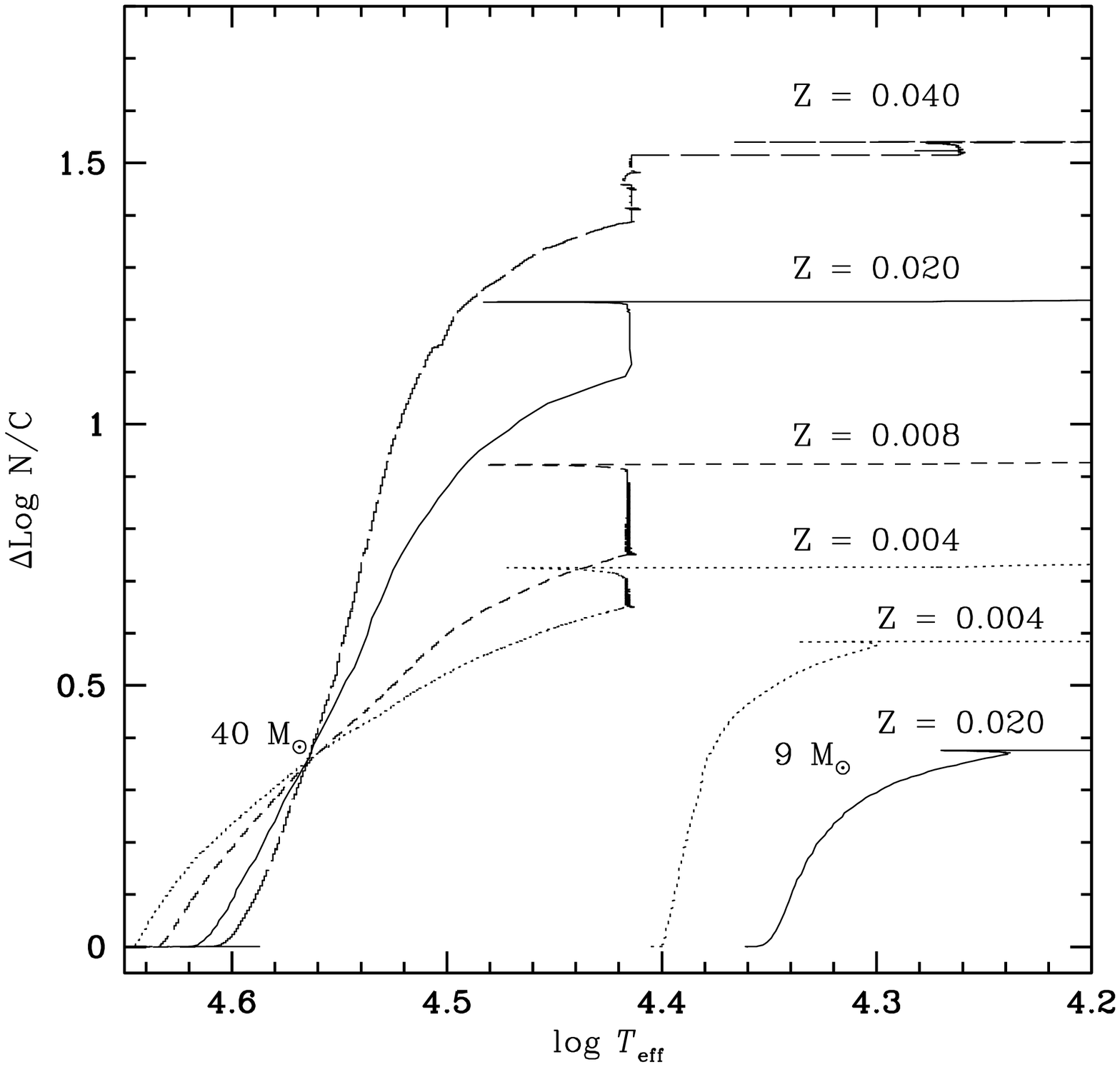}{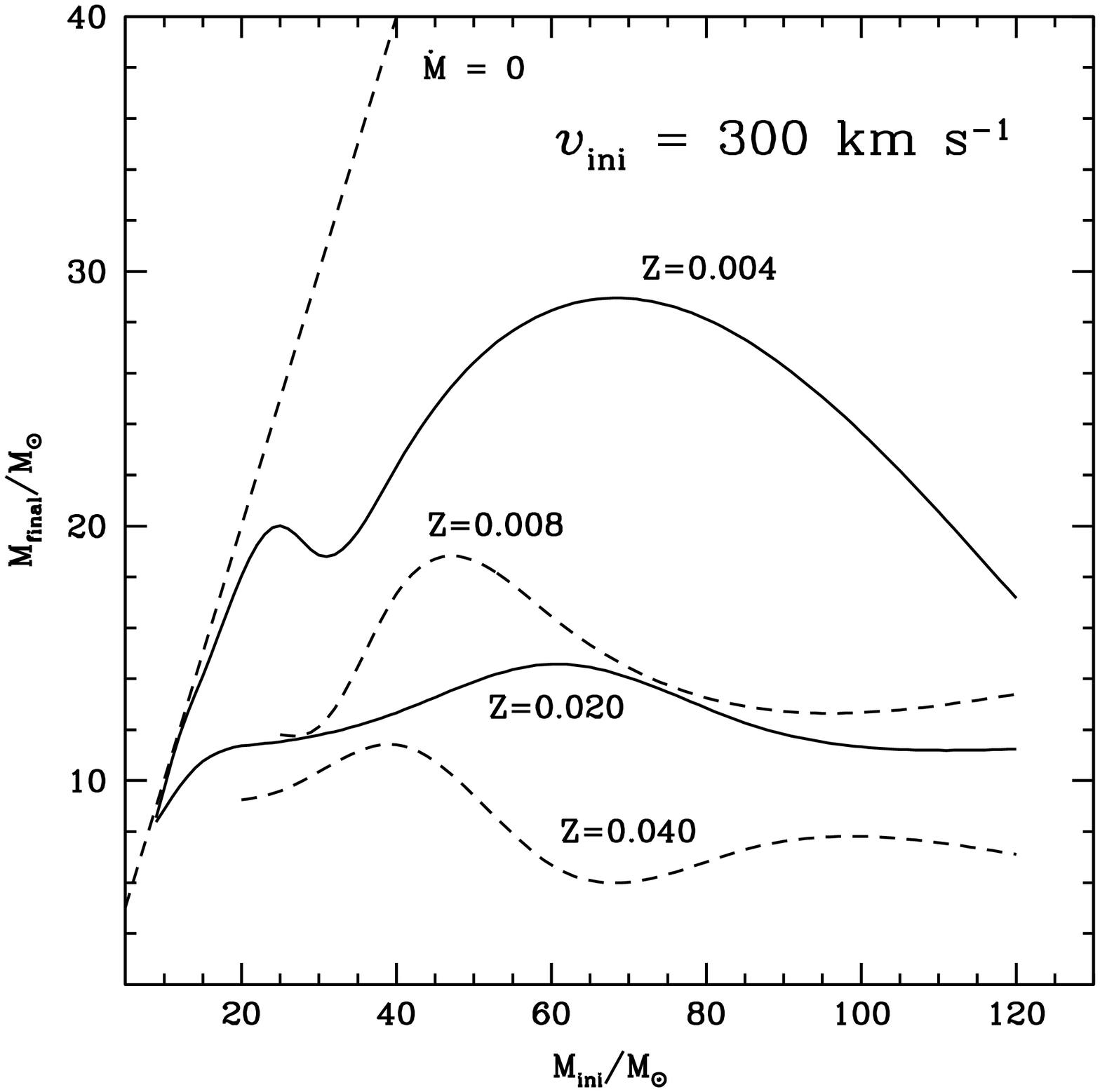}
\caption{Left: Relative enrichments in N/C for rotating stars of 9 and 40 $M_{\odot}$ at different $Z$
but with the same $v_{\mathrm{ini}}=300$ km/s.
Right:The final masses at the time of SN explosion as a function of mass and $Z$.}
\end{figure}%Fig.8

The variation of the H--surface content  
$X_{\mathrm{s}}$ as a function of  luminosity is a
constraining test (Fig.~4). The most massive stars go almost vertically down. 
The LBV stars, according to a recent collection of data \citep{Stothers2000}, are located 
between $\log L/L_{\odot}= 5.4$ and 6.2 with  $X_{\mathrm{s}}$ between 0.30 and 0.40.
The WN stars  are in almost the same range of $L$ while their $X_{\mathrm{s}}$
are between 0.45 and 0. This overlap supports the back and forth transition between the
WNL stars and the LBV. The further evolution from LBV leads them to WN stars, firstly 
WNL and then WNE. The WR lifetimes   as a function of  initial mass, rotation
  and  $Z$ have been calculated \citep{MMXI}.  The comparison with observations is considerably 
improved when rotation is accounted for.

Fig.~6 left shows the relative enrichments in $N/C$  for stars of 9 and 40 
$M_{\odot}$ at different $Z$. In the lower mass domain, the enrichments are
higher for stars with lower $Z$ since mixing is in general   stronger. This results
 from  the steeper internal
$\Omega$--gradients in lower $Z$ stars
\citep{MMVII}. For masses above 30 $M_{\odot}$, mass loss dominates the evolution
and as the $\dot{M}$--rates are larger at higher $Z$, the enrichments
are also larger at higher $Z$.  

\section{Final stages in relation with $\gamma$-ray bursts (GRBs)}

  \begin{figure}[!ht]
\plottwo{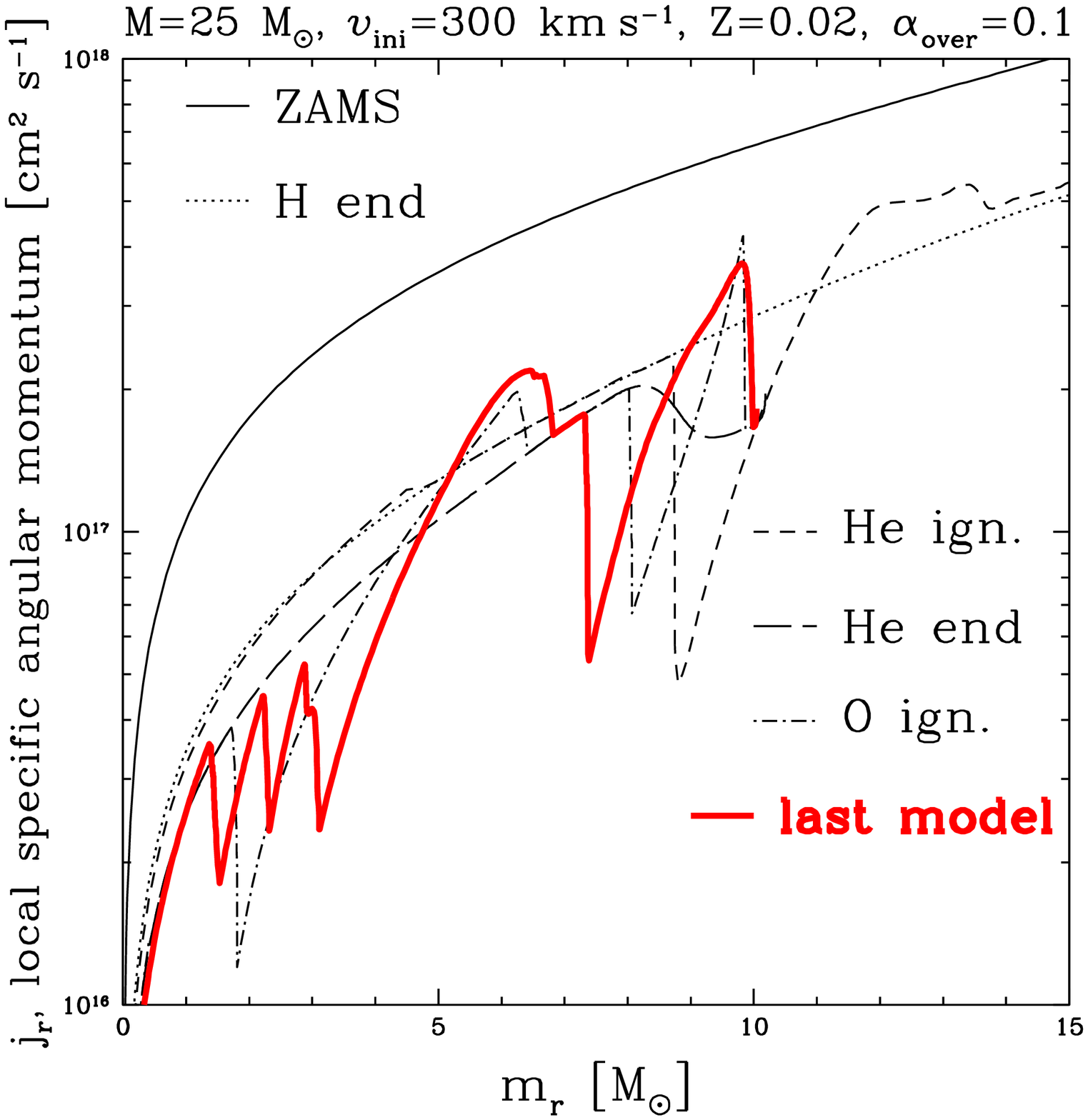}{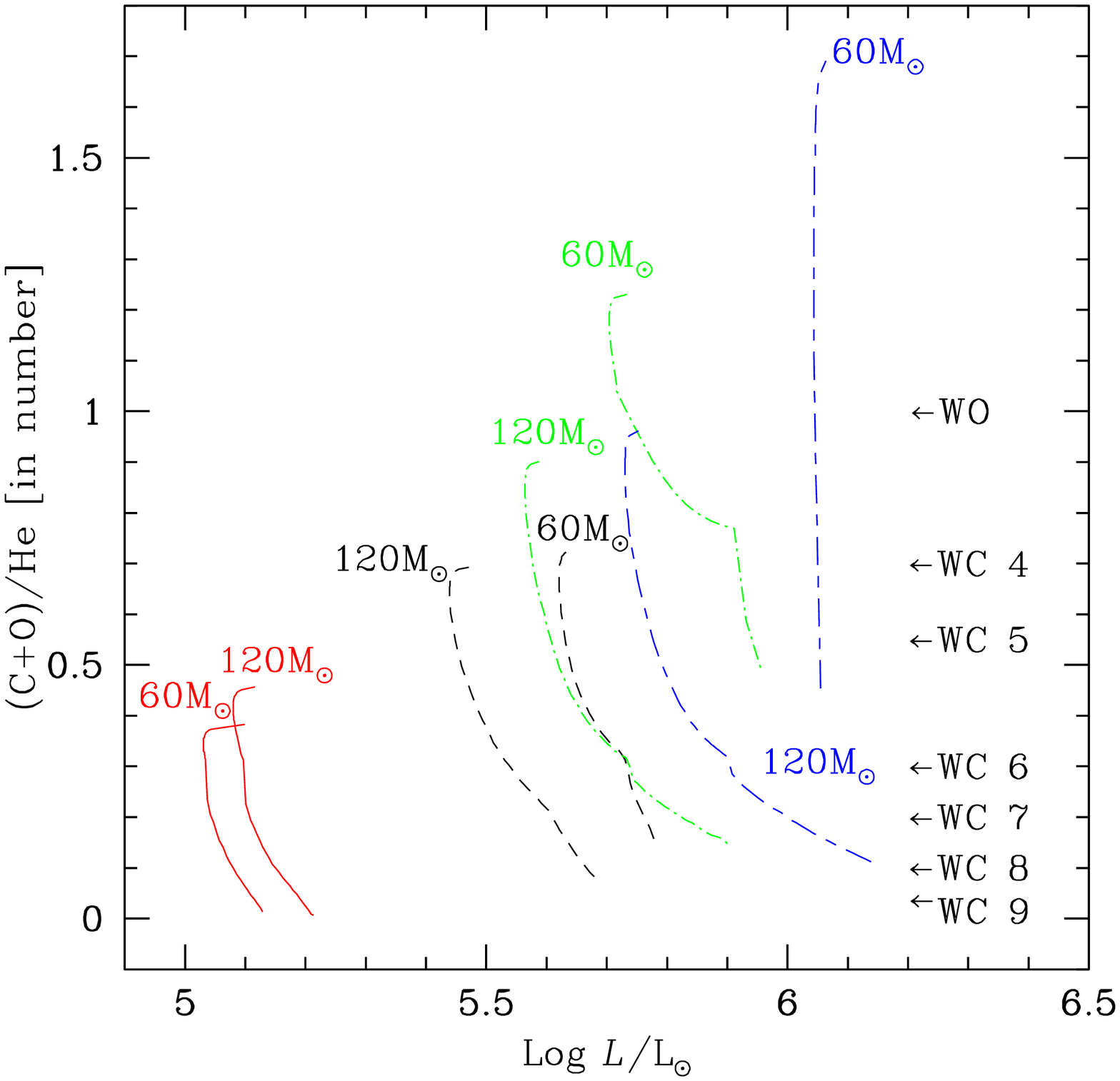}
\caption{Left: Distribution of the specific angular momentum in a model with an initial mass of
25 $M_{\odot}$. The distribution is shown at various evolutionary stages \citep{HMMXII}.
The distribution in the last model coincide well with results from 
Heger et al. \citep{HLW00,HWLS03}. Right: Evolution of the ratios (C+O)/He as a function of the luminosity at the surface of 60 and 120 $M_\odot$ rotating models for various initial 
  metallicities (see text). Long--short dashed curves show the evolution of $Z$ = 0.004 models, dashed--dotted curves, short--dashed curves
and continuous lines show the evolutions for $Z$ = 0.008, 0.020 and 0.040 respectively.
The correspondence between the (C+O)/He ratios and the different WC subtypes 
\citep{SmithM91} is indicated on the right.}
\end{figure}

As a result of mass loss, the stellar masses  decrease very much. Fig.~6 right
shows the final masses at the time of supernova explosion as a function of the initial masses 
and $Z$. For solar $Z$ or higher, all stars  with initial masses 
above 20 $M_{\odot}$ finish their life with 
masses of about 10 $M_{\odot}$. At lower $Z$, the final masses are higher.
It should not be concluded that  for $Z=0$ the final masses are equal to
the initial ones, because such stars are likely to spend a fraction of their 
MS lifetime at break--up and loose mass (see Meynet, this conference). The different final
masses lead to different types of supernovae.

Fig.~7 left shows the distribution of the specific angular momentum $j$ at various 
stages of an  initial  25 $M_{\odot}$ star. The final distribution
of $j$ is essentially shaped during MS evolution. Thus, the 
treatment of the transport of angular momentum in the MS phase is an essential aspect.
The various models \citep{HLW00,HWLS03,HMMXII} lead to 
rather  similar  distributions, despite
their  differences in  input physics.  A specific angular momentum above
 $10^{16}$ cm$^2$ s$^{-1}$ is necessary 
for the collapsar model to work  \citep{HLW00,HWLS03} 
and  account for the (GRB). These models fulfill 
this condition, however a lot of angular momentum has to be lost in the explosions
in order to account for the rotation rate of pulsars.

\begin{figure}[!ht]
\plottwo{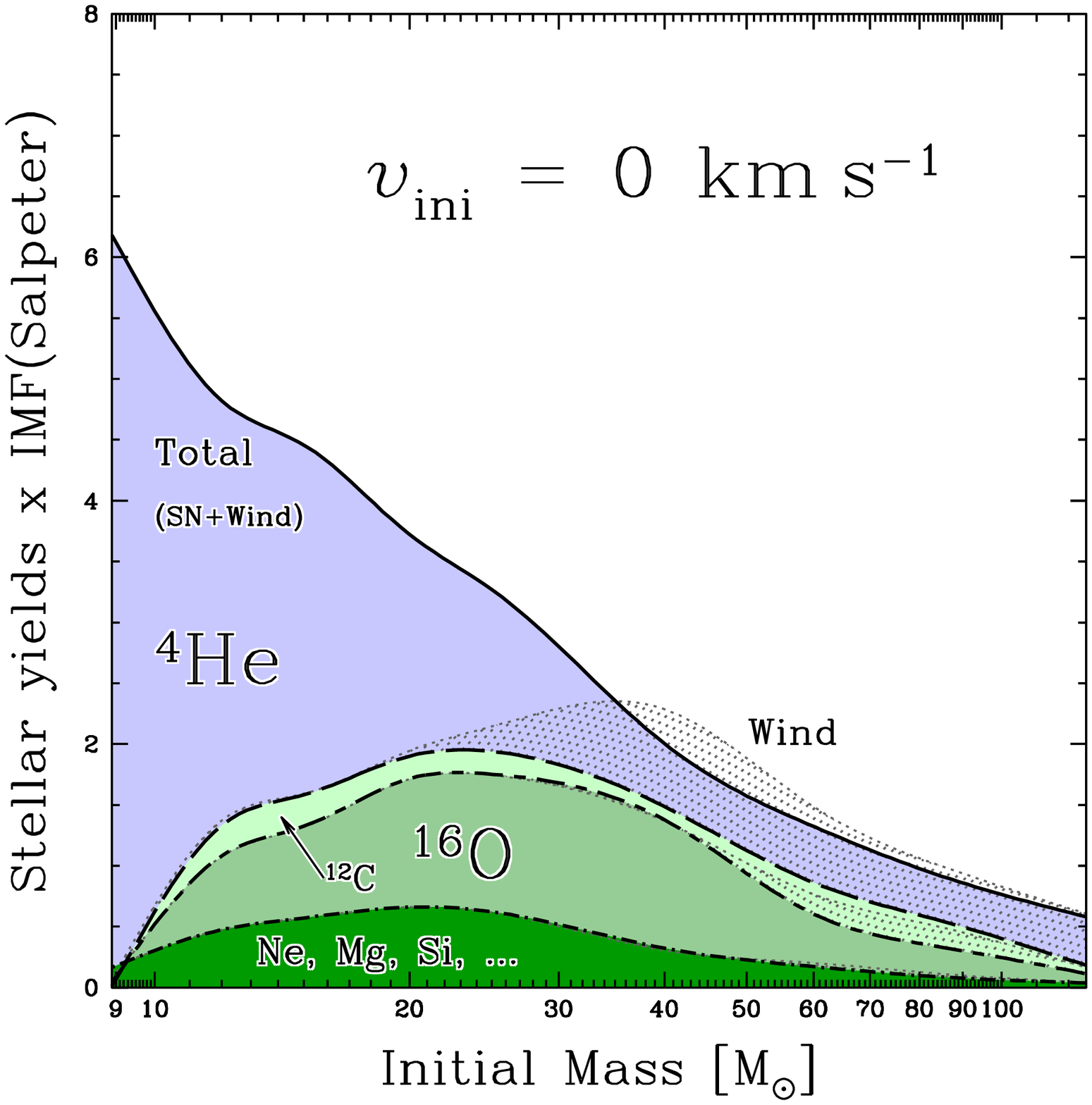}{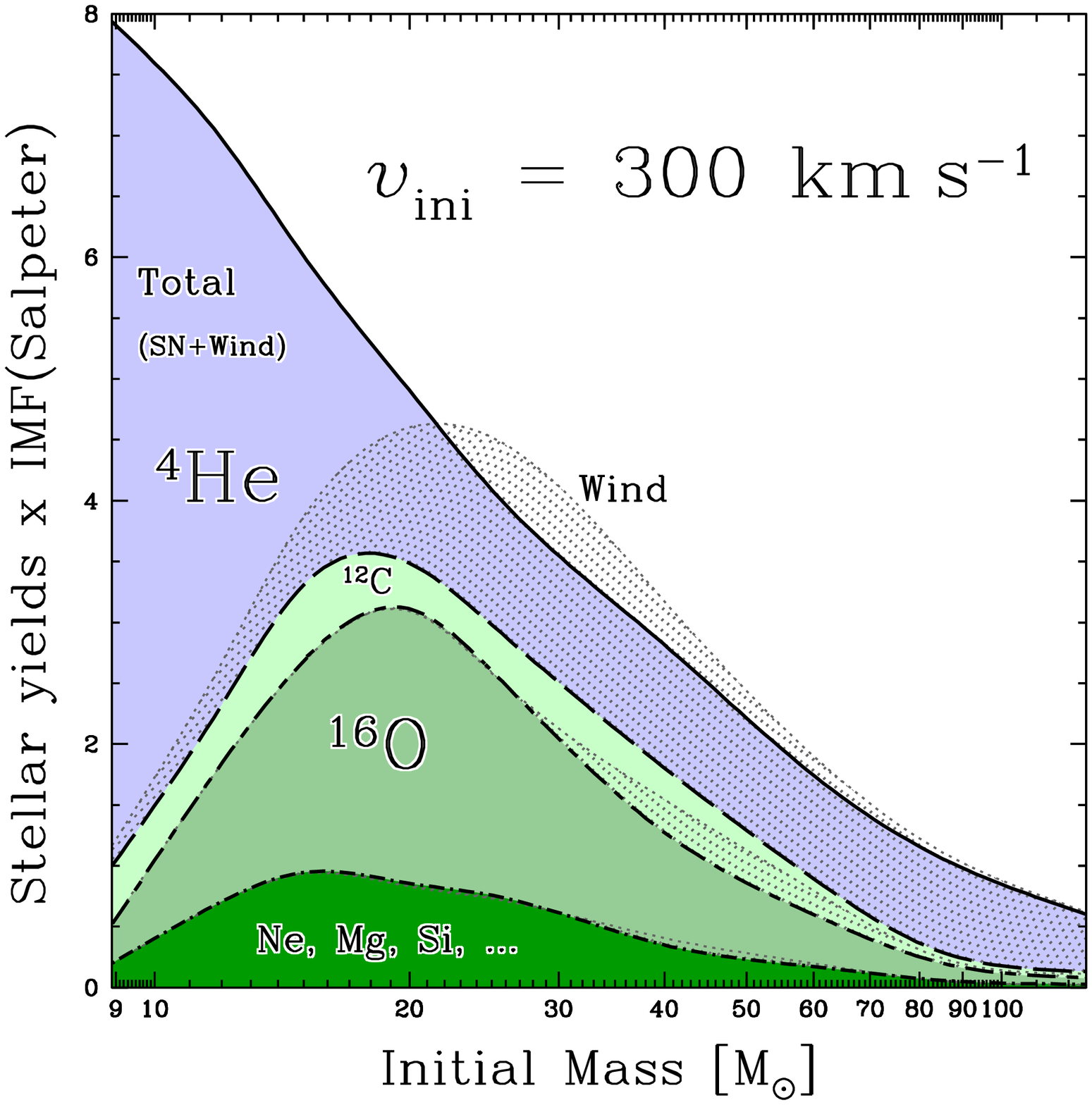}
\caption{Left: the yields x IMF for models without rotation. Right: the yields x IMF
for models with $v_{\mathrm{ini}}= 300$ km/s \citep{HMMXII}.}
\end{figure}

The association of GRB with hypernovae of the class of  SNIc
 is supported by several observations \citep{Mazzali03,Pods04}. SNIc result from the 
explosion of a star without H and with little or no He. 
This corresponds to a rare category of WR stars:
 the so--called WO stars.  WO stars ares subsequent to  WC3 objects, they 
show the products of He-burning, with an excess of C+O with respect to He and
 O $>$ C. WO stars are rare and result from the 
evolution of stars  with M $\geq$ 60 $M_{\odot}$ 
{\emph{at low metallicity only}} \citep{SmithM91}. These authors show that early WC 
types and WO stars are found in lower $Z$ regions. This may seem surprising.
The reason is \citep{SmithM91}: at high $Z$, mass loss is high, thus when the products
of the 3$\alpha$ reaction appear at the surface, they are in an early stage of nuclear processing,
i.e. with a low $C+O/He$ ratio (see Fig.~7 right). This corresponds to  late WC stars.
At low $Z$, the products of the 3$\alpha$ reaction rarely appear at the stellar surface
and if they do it (e.g. in case of high rotation) this  occurs very late in the evolution,
i.e. when $C+O/He$  is high and   this  is a WO star 
\citep{SmithM91}. Thus, the rare WO stars may be  the progenitors of SNIc and GRB.

Nucleosynthesis is also influenced by rotation \citep{HMMXII}. Below 30 $M_{\odot}$,
 the $\alpha$--elements are enhanced due to the larger cores. Above  30 $M_{\odot}$,
 mass loss is  the dominant effect and more He is ejected before being further processed.
 Fig.~8 shows the yields multiplied by the initial mass function.
 Due to the weighting by the  IMF, the production of oxygen and
 of $\alpha$--elements  is globally enhanced, while the effect on the He--production in massive stars
 is rather small.

\end{document}